\documentclass[reprint,
superscriptaddress,
amsmath,amssymb,
aps,
prl
]{revtex4-2}

\usepackage{graphicx}% Include figure files
\usepackage{dcolumn}% Align table columns on decimal point
\usepackage{bm}% bold math
\usepackage{amssymb}
\usepackage{breqn}
\usepackage{xcolor}
\usepackage{amsmath}
\usepackage{float}
\usepackage{tikz}
\usepackage{notoccite}

\begin{document}

\title{
Quantum-Enhanced Stimulated Brillouin Scattering Spectroscopy and Imaging
}

\author{Tian Li}
\email{tian.li@tamu.edu}
%\homepage{T.L. and F.L. contributed equally to this work.}
\affiliation{
Institute for Quantum Science and Engineering$,$ Texas A\&M  University${,}$  College Station${,}$   TX  77843${,}$  USA
}
\affiliation{
Department of Biological and Agricultural Engineering${,}$  Texas A\&M  University${,}$  College Station${,}$ TX 77843${,}$  USA
}
\affiliation{
Department of Chemistry and Physics${,}$  The University of Tennessee at Chattanooga${,}$  Chattanooga${,}$ TN 37403${,}$  USA
}

\author{Fu Li}
\affiliation{
Institute for Quantum Science and Engineering$,$ Texas A\&M  University${,}$  College Station${,}$   TX  77843${,}$  USA
}
\affiliation{
Department of Physics and Astronomy${,}$  Texas  A\&M  University${,}$ College Station${,}$  TX  77843${,}$  USA
}

\author{Xinghua Liu}
\affiliation{
Institute for Quantum Science and Engineering$,$ Texas A\&M  University${,}$  College Station${,}$   TX  77843${,}$  USA
}
\affiliation{
Department of Physics and Astronomy${,}$  Texas  A\&M  University${,}$ College Station${,}$  TX  77843${,}$  USA
}

% \author{Marlan O. Scully}
% \affiliation{
% Institute for Quantum Science and Engineering$,$ Texas A\&M  University${,}$  College Station${,}$   TX  77843${,}$  USA
% }
% \affiliation{
% Department of Mechanical and Aerospace Engineering${,}$  Princeton  University${,}$  Princeton${,}$  NJ  08544${,}$ USA
% }
% \affiliation{
% Quantum Optics Laboratory${,}$  Baylor  Research and  Innovation Collaborative${,}$  Waco${,}$ TX  76704${,}$  USA
% }

\author{Vladislav V. Yakovlev}
\affiliation{
Department of Physics and Astronomy${,}$  Texas  A\&M  University${,}$ College Station${,}$  TX  77843${,}$  USA
}
\affiliation{
Department of Biomedical Engineering$,$ Texas A\&M  University${,}$  College Station${,}$   TX  77843${,}$  USA
}
\affiliation{
Department of Electrical and Computer Engineering$,$ Texas A\&M  University${,}$  College Station${,}$   TX  77843${,}$  USA
}

\author{Girish S. Agarwal}
\affiliation{
Institute for Quantum Science and Engineering$,$ Texas A\&M  University${,}$  College Station${,}$   TX  77843${,}$  USA
}
\affiliation{
Department of Biological and Agricultural Engineering${,}$  Texas A\&M  University${,}$  College Station${,}$ TX 77843${,}$  USA
}
\affiliation{
Department of Physics and Astronomy${,}$  Texas  A\&M  University${,}$ College Station${,}$  TX  77843${,}$  USA
}

%\date{\today}

\begin{abstract}

Brillouin microscopy is an emerging label-free imaging technique to assess local viscoelastic properties. Quantum-enhanced stimulated Brillouin scattering is demonstrated for the first time using low power continuous-wave lasers at 795~nm.  A signal to noise ratio enhancement of 3.4~dB is reported by using two-mode intensity-difference squeezed light generated with the four-wave mixing process in atomic rubidium vapor. The low optical power and the excitation wavelengths in the water transparency window has the potential to provide a powerful bio-imaging technique for probing mechanical properties of biological samples prone to phototoxicity and thermal effects. \textcolor{black}{The performance enhancement affordable through the use of quantum light may pave the way for significantly improved sensitivity that cannot be achieved classically.}
%, data acquisition rate and spatial resolution. 
The proposed new way of utilizing squeezed light for enhanced stimulated Brillouin scattering can be easily adapted for both spectroscopic and imaging applications in biology.

\end{abstract}

%\ocis{270.1670, 020.1670, 020.2070}
%OCIS:
%020 atomic physics
%020.1670   Coherent optical effects
%020.2070   Effects of collisions
%030 coherence
%030.1670   Coherent optical effects
%270 quantum optics
%270.1670
%300 spectroscopy
%300.6210 spectroscopy, atomic

%\pacs{42.50.Gy, 32.80.Qk, 42.65.-k}

% 34.50.-s  Scattering of atoms and molecules
%34.20.-b   Interatomic and intermolecular potentials and forces, potential energy surfaces for collisions
%32.80.Xx   Atomic properties and interactions with photons, Level crossing and optical pumping
% 95.30.Dr  Fundamental aspects of astrophysics, Atomic processes and interactions
% 32.30.Jc  Visible and ultraviolet spectra
% 34.20.Cf  Interatomic potentials and forces
% 34.50.Cx  Elastic; ultracold collisions
% 37.10.De  Atom cooling methods
% 32.60.+i  Zeeman and Stark effects
% 32.80.Qk  Coherent control of atomic interactions with photons
% 42.50.Ex  Optical implementations of quantum information processing and transfer
% 42.50.Gy  Effects of atomic coherence on propagation, absorption, and amplification of light; electromagnetically induced transparency and absorption
% 42.65.-k Nonlinear optics

\maketitle

%\section{Introduction}
Over the past decade, Brillouin scattering spectroscopy and microscopy has witnessed its renaissance providing solutions to fundamental problems and sparking new applications across multiple disciplines~\cite{kargar2021advances,2020,prevedel2019brillouin,palombo2019brillouin,PhysRevX.7.031015,Meng:16}. In the innermost part of those revolutionary advancements is the new ways of improving detection either through high-resolution spectrometer~\cite{scarcelli2008confocal} or through nonlinear optical excitation~\cite{remer2020high,ballmann2017impulsive,ballmann2015stimulated}. Brillouin scattering is an inelastic scattering of light by electrostrically or thermally excited acoustic waves (i.e., phonons). 
%The phenomenon has been known since the 1920's. due to the independent theoretical predictions by L. Brillouin and L. Mandelstam and first experimental observation by E. Gross (for more in depth review of the history and instrumentation development, see~\cite{kargar2021advances}). 
%\textcolor{black}{//Now, here we need a concept figure, explaining the effect of Brillouin scattering, what is measured (show spectrum and indicate Brillouin shift and linewidth), stimulated Brillouin scattering and the concept of squeezed light or ``quantum advantage".//} 
If a narrow linewidth ($<$~10 MHz) light source is used, both the red-shifted (Stokes) and blue-shifted (anti-Stokes) scattered light is detected giving rise to a Brillouin spectrum.
% which is dependent on the scattering angle and local viscoelastic properties that are embedded through the speed of ultrasound and its attenuation through the known relationships~\cite{NonlinearOptics}: $\Omega_{\text{B}}= 2n\omega(v/c)\text{sin} (\theta/2)$ and $\Gamma_{\text{B}}= (2\eta_{\text{s}}+\eta_{\text{d}})q^2/\rho_0$, where $\Omega_{\text{B}}$ and $\Gamma_{\text{B}}$ are the Brillouin frequency shift and Brillouin linewidth respectively, $n$ is the index of refraction of the medium, $\omega$ is the frequency of the incident light, $v$ is the speed of sound in the medium, $c$ is the speed of light in vacuum, $\theta$ is the angle between the incoming and backscatterd light, $q$ is the phonon wave-number, $\rho_0$ is the average density of the medium, and $\eta_{\text{s}}$ and $\eta_{\text{d}}$ are the shear and dilational viscocity coefficients respectively.
By measuring both the frequency shift and the linewidth of the spectrum, the complex viscoelastic modulus of the sample can be assessed in a single spectroscopic measurement~\cite{bottani2018brillouin}. 
%Traditional applications of Brillouin scattering involved characterization of mechanical properties of solid-state materials, optical fibers and gems~\cite{bottani2018brillouin} and involved spontaneous Brillouin scattering, which is based on detecting the spectrum of scattered photons. 
Relatively recently, biological applications of Brillouin spectroscopy became a subject of interest~\cite{vaughan1980brillouin}. Recent years brought a deeper understanding of microscopic biomechanics as one of the key governing factors in biological development and diseases such as cancer progression~\cite{bausch2006bottom}. Brillouin scattering spectroscopy offers a non-contact, label free method, it is therefore very suited for measurements of biomechanical properties that would be difficult to measure with other methods~\cite{ballmann2019nonlinear,bao2003cell}.

\begin{figure}[t]
    \begin{center}
    \includegraphics[width=1\linewidth]{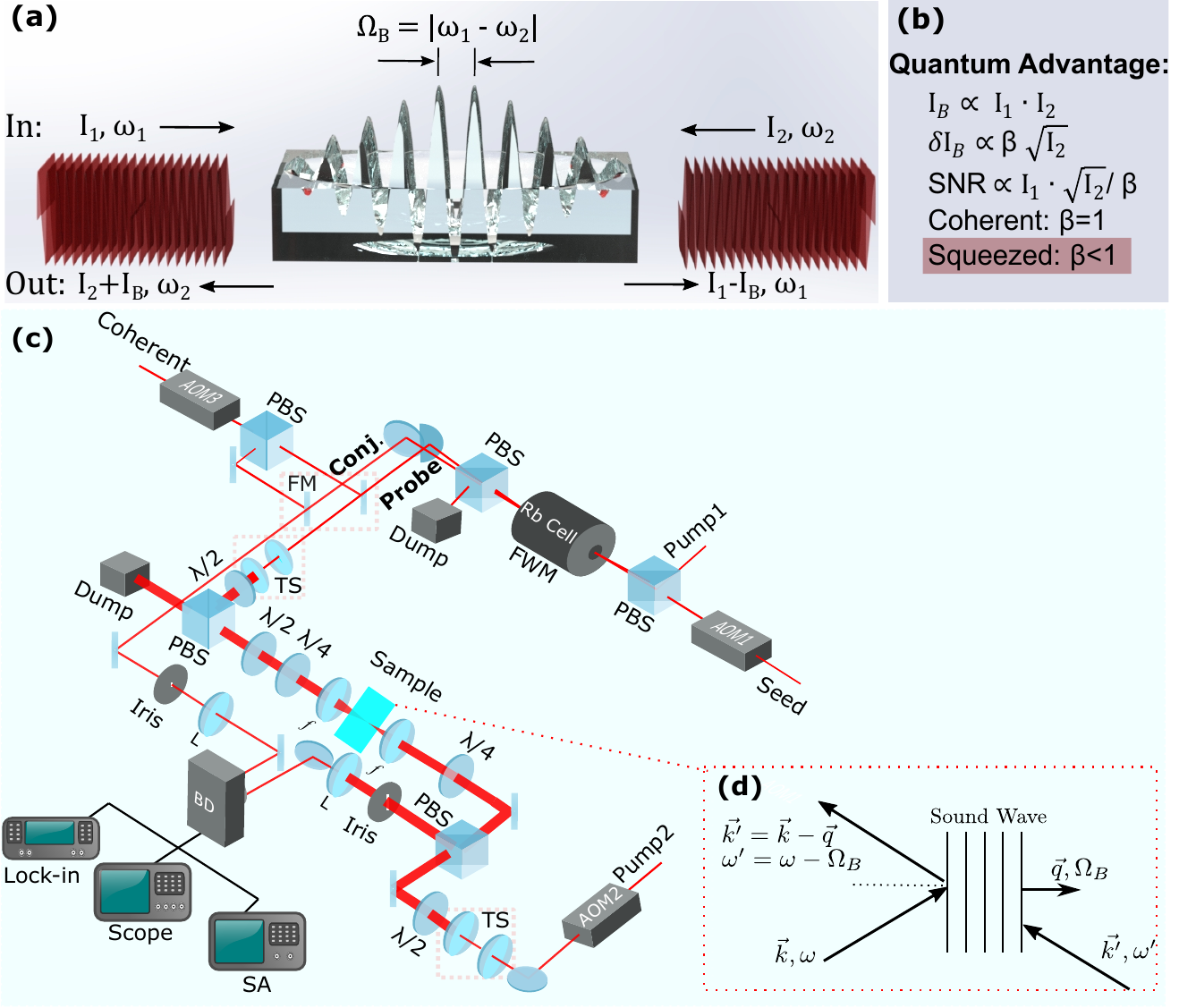}
    \caption{
     \textcolor{black}{(a) Conceptual diagram of the SBS process. The stimulated Brillouin gain $\text{I}_{\text{B}}$ is what we measure in this work. (b) Principle of quantum advantage for the measurement of $\text{I}_{\text{B}}$.} (c) Experimental setup. 
     %See text for a detailed description.
%in which a seeded $^{85}$Rb vapor cell produces strong quantum-correlated twin beams (``Probe" and ``Conj.") via FWM. The twin beams are separated from the pump (``Pump 1") by a $\sim$~$2\times10^5$~$:1$ Glan-Laser polarizer after the cell. The probe beam and another pump beam (``Pump 2") form a geometry for the SBS process in water (see text for details) while the conjugate beam serves as a reference, before they are focused onto a balanced detector (BD). Two flip mirrors (FM) are used to steer in two coherent beams to form a classical version of the whole setup. The three AOMs in the beam paths are used to amplitude-modulate individual beams. 
L: lens, FM: flip mirror, TS: telescope, PBS: polarizing beam splitter, BD: balanced detector, SA: RF spectrum analyser.
%(b) Level structure of the D1 transition of $^{85}$Rb atom. The optical transitions are arranged in a double~-~$\Lambda$ configuration, where $\nu_p$, $\nu_c$ and $\nu_1$ stand for probe, conjugate and pump frequencies, respectively, fulfilling $\nu_p$ +  $\nu_c$ =  $2\nu_1$ and $\nu_c - \nu_p = 2\nu_{HF}$. The width of the excited state in the level diagram represents the Doppler broadened line. $\Delta$ is the one-photon detuning. $\nu_{\text{HF}}$ is the hyperfine splitting in the electronic ground state of $^{85}$Rb.
 (d) Phase-matching diagram for the SBS process~\cite{NonlinearOptics}. The wave-vectors and frequencies for the pump, probe and sound wave are denoted by ($\vec{k}$, $\omega$), ($\vec{k^{\prime}}$, $\omega^{\prime}$) and ($\vec{q}$, $\Omega_B$) respectively.
        \label{Setup1}}
    \end{center}
\end{figure}

With all the advantages of Brillouin spectroscopy being able to provide unique information in a remote and noninvasive way, there is still a tremendous amount of remaining challenges to improve the accuracy and acquisition speed of such measurements, in order to observe fast dynamic processes and to image large scale objects with microscopic spatial resolution.
%Optical coherence~\cite{petrov2007comparison} provides an exquisite tool to enhance the signal, and its advantages for Brillouin spectroscopy and imaging have been successfully demonstrated~\cite{ballmann2015stimulated,ballmann2017impulsive,remer2020high}.
\textcolor{black}{To improve the acquisition speed and spatial resolution, and to reduce the elastic scattering background, stimulated Brillouin scattering (SBS) was proposed and was first observed by Chiao \textit{et al.}~\cite{PhysRevLett.12.592}. Since then SBS has been demonstrated to enable much faster acquisition times and high specificity to local viscoelastic properties of biological samples~\cite{ballmann2015stimulated,remer2020high}. 
%SBS provides a unique way of enhancing the signal, which was first observed by Chiao \textit{et al.}~\cite{PhysRevLett.12.592}.  
As shown by the diagram in Fig.~\ref{Setup1}(a), in the process of SBS, counter propagating continuous-wave (CW) pump and probe beams at frequencies $\omega_1$ and $\omega_2$ overlap in the sample to efficiently interact with a longitudinal acoustic phonon of frequency $\Omega_{\text{B}}$. When $\omega_2$ is scanned around the Stokes frequency ($\omega_1-\Omega_{\text{B}}$), the probe intensity $\text{I}_2$ at $\omega_2$ experiences a stimulated Brillouin gain ($+\text{I}_{\text{B}}$) via wave resonance, where the pump intensity $\text{I}_1$ at $\omega_1$ shows a stimulated Brillouin loss ($-\text{I}_{\text{B}}$). The opposite occurs when $\omega_2$ is scanned around the anti-Stokes frequency ($\omega_1-\Omega_{\text{B}}$). Thus, the stimulated Brillouin gain and loss enable spectral measurements that are free of elastic background.} The efficiency of phonon generation in SBS 
%is proportional to the intensity of the stronger one of these two beams, and hence the Brillouin scattering 
could be orders of magnitude stronger than in the scenario of spontaneous Brillouin scattering~\cite{ballmann2015stimulated}. High scattered signal magnitude translates into better signal-to-noise ratio (SNR)
%of stimulated versus the spontaneous techniques, 
and consequently faster acquisition times. With all the benefits provided by SBS, however, there are also two main technical challenges associate with it: 1) locking laser frequency to an external reference, such as a cavity or an absorption line, is required to reduce temporal drifts of laser frequency~\cite{Meng:16,Coker:18}, and 2) alignment of the two counter-propagating laser beams must be very precise for a significant spatial overlap of their individual focal regions (i.e., Rayleigh ranges).
As a nonlinear optical technique, SBS benefits from the higher excitation intensity, which can also induce phototoxicity and/or thermal damage to biological samples of interest. Clearly, there is a tremendous need to improve SBS detection for low-power-light applications, and recent advancements in LIGO demonstrate a path to improve the detection limit using squeezed light spectroscopy~\cite{yu2020quantum,Unruh}. Since photon shot-noise is a fundamental limit for optical detection, it is thus intuitive to deploy the strategy utilizing quantum light to beat this limitation. In fact, applications of quantum light in the context of Raman spectroscopy have been discussed very recently~\cite{casacio2021quantum,de2020quantum,PhysRevResearch.3.043029,RevModPhys.88.045008}. In this letter, we demonstrate the very first quantum-enhanced SBS spectroscopy and imaging that is capable of improving the measurement SNR by 3.4~dB \textcolor{black}{(or improving \textit{sensitivity} by  1.7~dB~\cite{PhysRevApplied.15.044030})} when compared to classical excitation with the same incident light intensity. \textcolor{black}{Our scheme adopts the standard `modulation-demodulation' type of approach (so that weak signal could appear within the spectrum range where the noise level is, or close to, shot-noise limited) that has been employed in some state-of-the-art demonstrations~\cite{remer2020high,ballmann2015stimulated}, thus the absolute signal size here is comparable to theirs, only because our quantum light reduces the noise level, therefore the SNR is enhanced with respect to classical techniques. The principle of this quantum enhancement is shown in Fig.~\ref{Setup1}(b). The physical quantity that we measure throughout this work is the stimulated Brillouin gain $\text{I}_{\text{B}}$. If the uncertainty of this measurement $\delta\text{I}_{\text{B}}$ can be quantum-mechanically `squeezed ($\beta<1$)' so that it is below the shot-noise level ($\beta=1$), then the measurement SNR can be therefore subsequently improved by a factor of $1/\beta$.}

%\textcolor{black}{//Now spend several paragraphs describing what is squeezed light, how it generates the quantum advantage, what limits it, etc. Essentially provide some theoretical background, explaining what, why and how. Since we are not developing here any theory, at the end of it, we just need to clearly summarize what is needed to demonstrate this effect, how we will demonstrate it and what is the impact.//} 

The quantum state of light used in this scheme is a two-mode intensity-difference squeezed light generated with the four-wave mixing (FWM) process in an atomic $^{85}$Rb vapor cell, which has proven to be a great platform for quantum sensing applications~\cite{dorfman2021multidimensional,PhysRevApplied.15.044030,prajapati2021quantum,Dowran:18,Anderson:17,Li:17,Pooser:15}. %The atomic medium possesses a large third-order electric susceptibility $\chi^{(3)}$, and when appropriately chosen laser light (at 795~nm, near $^{85}$Rb $D_1$ line) `seeds' the medium, `twin beams', also known as the `probe' and `conjugate' beams, are produced.
Major advantages of this FWM-based quantum light generation scheme are strong intensity-difference squeezing (greater than 6~dB)
%include an ultra-high entangled photon-pair flux up to~$10^{16}$ photons/s, which is a few orders of magnitude higher than fluxes produced by spontaneous parametric down-converters~\cite{Varnavski:2017rp,Jechow:08}, 
and narrow-band twin beams ($\sim10$~MHz)~\cite{PhysRevResearch.3.033095,Clark:2014vf,Glasser2012a}, which is extremely beneficial for the intended SBS experiment, where the spectral width of the light source must be well below the Brillouin linewidth, which is typically a few hundreds of MHz. The SNR for the twin beams, with signal defined as the difference of photon numbers in the twin beams, is better than that for coherent beams by a factor of $\text{cosh}2r$, where $r$ is the well known squeezing parameter used to characterize the two-mode squeezed state~\cite{NonlinearOptics}. This improvement in SNR consequently translates to the quantum advantage in the SBS spectroscopy (See Eq.~(10) in Ref.~\cite{SM}).  
%The total excitation power used in our scheme can be as little as $\sim 8$~mW, it can be therefore easily adapted for specific applications to probe delicate samples. The low optical power together with the excitation wavelengths in the water transparency window are highly desirable for biological applications as they provide a unique tool by which mechanical properties of biological samples can be characterized without introducing significant disturbance.  

The schematic of our experimental setup is shown in Fig.~\ref{Setup1}(c). Two quantum-correlated beams of light, i.e., the ``probe'' and ``conjugate'' beams, are produced with the FWM process in the $^{85}$Rb vapor cell. After the cell, the probe beam is overlapped with a counter-propagating laser beam (shown in Fig.~\ref{Setup1}(c) as ``Pump 2") at a homemade sample holder filled with distilled water, to form a phase-matching geometry for the SBS process depicted in Fig.~\ref{Setup1}(d). The conjugate beam serves as a reference, and two flip mirrors (FM) are used for the introduction of two coherent beams so that the whole setup can be converted to a classical version. 
%The pumps and probe beams are amplitude-modulated by three acousto-optic modulators (AOMs) at 300~KHz (AOM1) and 400~KHz (AOM2\&3) respectively. 
The water SBS gain is expected to appear at 700~KHz \textcolor{black}{(sum frequency of the amplitude modulations on the two involving beams~\cite{SM})} where the two-mode squeezing is expected to be the best~\cite{PhysRevApplied.15.044030,doi:10.1063/5.0010909}. Both pump lasers are locked to external cavities so that the relative frequency between them can be scanned \textcolor{black}{with 40~MHz spectral resolution.} \textcolor{black}{Also note that, the SBS gain is measured by a customized balanced detector, which subtracts away common-mode technical noise of the two input beams to better than 25~dB, so that noise level at 700~KHz (where signal appears) is shot-noise limited.} Other experimental details can be found in Ref.~\cite{SM}.

\begin{figure}[t]
    %\begin{center}
    \includegraphics[width=1\linewidth]{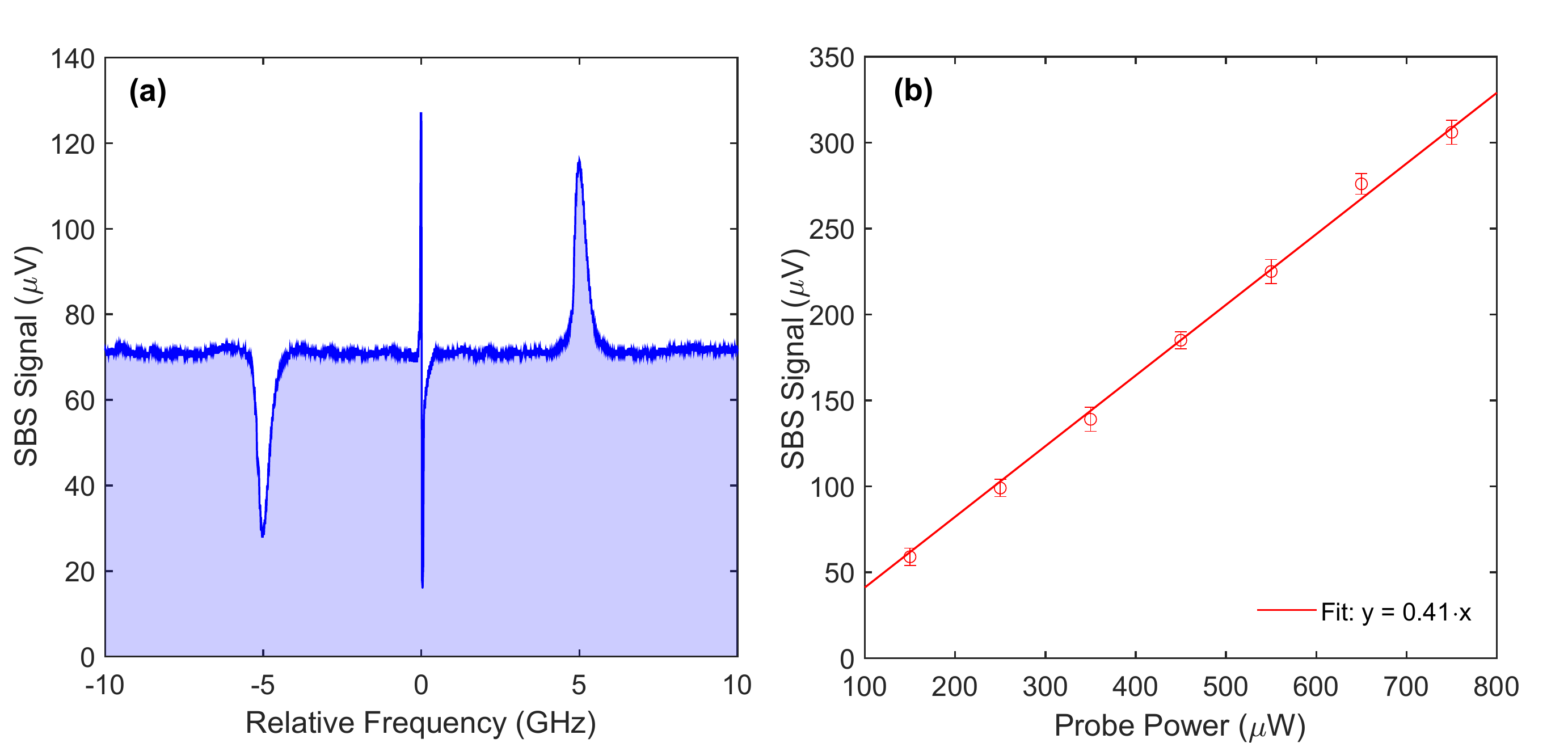}
    \caption{(a) SBS spectrum of water obtained by a lock-in amplifier.
    %The loss dip and gain peak of the SBS process are on the left and right of origin respectively, while the center feature is from absorptive stimulated Rayleigh scattering. The Brillouin frequency shift is measured to be $\Omega_B/2\pi$ = 5.01~GHz with a Brillouin linewidth of $\Gamma/2\pi$ = 292~MHz. 
    (b) Linear dependency can be clearly observed between the water SBS signal (at the peak of gain) and the optical power of a coherent probe.} 
    %\end{center}
    \label{LockIn}
\end{figure}

We start with classically characterizing the water SBS gain. Instead of using the squeezed twin beams, \textcolor{black}{we use \textit{a ``balanced coherent detection''} (by flipping up the two flip mirrors shown in Fig.~\ref{Setup1}(c)) for this classical measurement since our quantum light is in a \textit{``two-mode"} squeezed state where in addition to the probe beam, we always have a quantum-correlated reference/conjugate beam, and the squeezing resides in the \textit{``intensity-difference"} of the two involving beams.} Figure~\ref{LockIn}(a) shows a SBS spectrum of distilled $\text{H}_2\text{O}$ (T = 21~$^\circ$C) from a lock-in amplifier (with 300~ms time constant). The coherent beam in the pathway of the probe beam (i.e., the coherent probe) is locked while the pump beam of the SBS process (``Pump 2" in Fig.~\ref{Setup1}(c)) is scanned with 0.02~Hz scan frequency. The optical powers of the coherent and pump beams at the sample are 300~$\mu$W and 36~mW respectively. From Fig.~\ref{LockIn}(a) the Brillouin shift and linewidth are measured to be $\Omega_B/2\pi = 5.01 \textcolor{black}{\pm0.17}$~GHz and $\Gamma_B/2\pi = 292 \textcolor{black}{\pm27}$~MHz, which are in good agreement with previous experiments~\cite{ballmann2015stimulated}. The dip on the left (at $\sim-5$~GHz) and peak on the right (at $\sim5$~GHz) of zero are the stimulated Brillouin loss and gain peaks respectively. The center feature is caused by absorptive stimulated Rayleigh scattering. When we change the power of the coherent probe from 150~$\mu$W to 750~$\mu$W while keeping pump power at 36~mW, as shown in Fig.~\ref{LockIn}(b), we clearly observe the expected linear dependency between the SBS gain and the optical power of the coherent probe~\cite{ballmann2015stimulated}. Estimation of the SBS gain magnitude can be found in Ref.~\cite{SM}.

\begin{figure}[]
    %\begin{center}
    \includegraphics[width=1.00\linewidth]{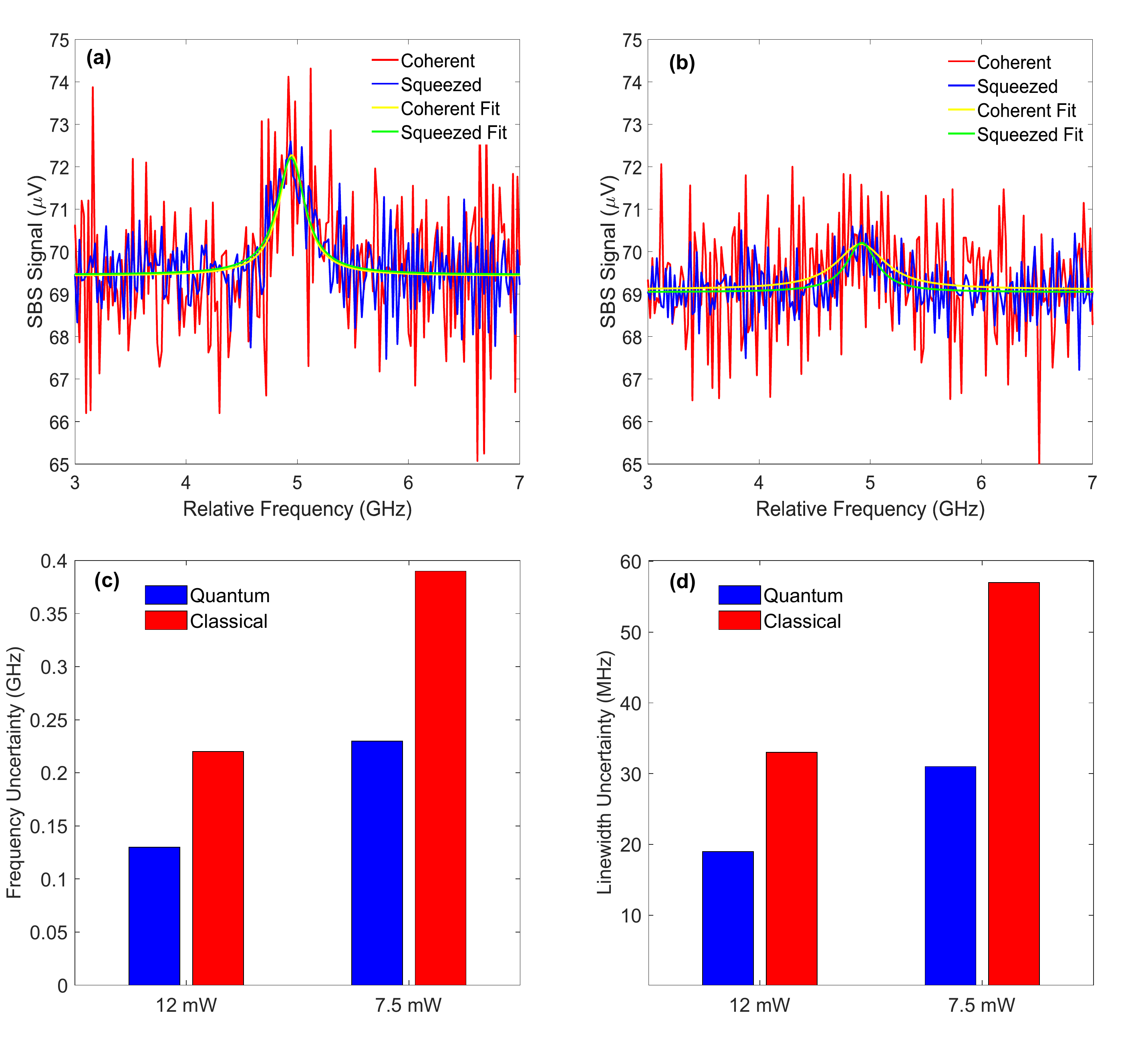}
    \caption{\textcolor{black}{Demonstration of quantum-enhanced estimations of the Brillouin scattering parameters (i.e., Brillouin shift and linewidth) of water, using probe power of 750~$\mu$W and pump power of (a) 12~mW and (b) 7.5~mW. The water SBS spectra were acquired by a lock-in amplifier. The red and blue spectra  correspond to the configurations where the probe beam is in a coherent state and in a two-mode squeezed state respectively. 
    %with -~3.40~dB noise reduction below the shot-noise level. 
    The yellow and green curves are the Lorentzian fits for the red (coherent) and blue (squeezed) spectra respectively. The bar plots in (c) and (d) represent the uncertainties of estimations of the Brillouin parameters for both the quantum and classical configurations.}
        \label{2Pumps_LockIn}}
    %\end{center}
\end{figure}

Having characterized the classical SBS process in water, in the following we demonstrate the quantum-enhanced water SBS spectra. To clearly demonstrate quantum-improved performance beyond the classical approach, we conducted the experiment both with the probe beam in a coherent state and in the two-mode squeezed state. The experimental scheme can be easily swapped between the two operations simply by flipping up and down the two flip mirrors depicted in Fig.~\ref{Setup1}(c). \textcolor{black}{We first use the lock-in amplifier to acquire Brillouin spectra of water to assess frequency shift $\Omega_B$ and linewidth $\Gamma_B$ at different pump powers under both the quantum and classical configurations. Note that the Brillouin scattering parameters, $\Omega_B$ and $\Gamma_B$, are not optical power dependent. We present quantum-enhanced estimations of the Brillouin scattering parameters in Fig.~\ref{2Pumps_LockIn}. Figures~\ref{2Pumps_LockIn}(a) and~\ref{2Pumps_LockIn}(b) show typical Brillouin spectra of water at room temperature for the pump powers of 12~mW and 7.5~mW respectively and show both the coherent light (red trace) and squeezed light (blue trace) experimental data with their respectful Lorentzian fits. The probe power for both cases was kept at 750~$\mu$W. We took 20 spectra each for the classical and quantum cases with the two pump powers, and fit these spectra with Lorentzian curves. We can clearly see the amplitude of fluctuations/noise for the coherent case (red trace) is much larger (greater than 2 times) than that for the squeezed case (blue trace). In Fig.~\ref{2Pumps_LockIn}(a), i.e., when the pump power was 12~mW, the fit parameters in the coherent case are: Brillouin shift $\Omega_B/2\pi=5.07\pm0.22$~GHz, and linewidth $\Gamma_B/2\pi=269\pm33$~MHz; whereas in the squeezed case, these parameters are: $\Omega_B/2\pi=5.01\pm0.13$~GHz, and $\Gamma_B/2\pi = 277\pm19$~MHz. Error bars in the fitting parameters represent one standard deviation. Examples of these fits are shown as the yellow and green Lorentzian curves for the coherent and squeezed cases respectively.  When the pump power was 7.5~mW, shown in Fig.~\ref{2Pumps_LockIn}(b), the Brillouin gain profile of water is almost completely buried by the fluctuations/noise in the coherent case, which renders the fit parameters with huge error bars, here the fit parameters are: $\Omega_B/2\pi=5.14\pm0.39$~GHz, and $\Gamma_B/2\pi=521\pm57$~MHz. We can clearly see a broadened linewidth from the coherent fit depicted by the yellow Lorentzian curve, which implies that the fit is even not physically meaningful under this condition, as $\Gamma_B/2\pi=521$~MHz is very far from the theoretical value of 289~MHz~\cite{ballmann2015stimulated}. Whereas in the squeezed case, these fit parameters are still acceptable, with $\Omega_B/2\pi=4.91\pm0.23$~GHz, and $\Gamma_B/2\pi=314\pm31$~MHz. By analyzing these data, we see that the improvement of SNR through the reduction of noise (by a factor of 2) in quantum measurements translates to a corresponding improvement of accuracy of the peak position and linewidth measurements by approximately the same amount (a factor of 2) for absolutely identical conditions of measurement (i.e. pump power, focusing spot-size, detectors used, acquisition times, etc.). From the bar plots in Figs. ~\ref{2Pumps_LockIn}(c) and~\ref{2Pumps_LockIn}(d), we can clearly see the improvement for quantum-enhanced measurements of the Brillouin parameters as compared to that for classical measurements.}

%Comparing with the expected Brillouin shift and linewidth of water at room temperature: $\Omega_B/2\pi=4.97$~GHz and $\Gamma_B/2\pi=289$~MHz~\cite{ballmann2015stimulated}, we see that the squeezed case always yields a much more accurate estimation.

\begin{figure}[]
    %\begin{center}
    \includegraphics[width=1.00\linewidth]{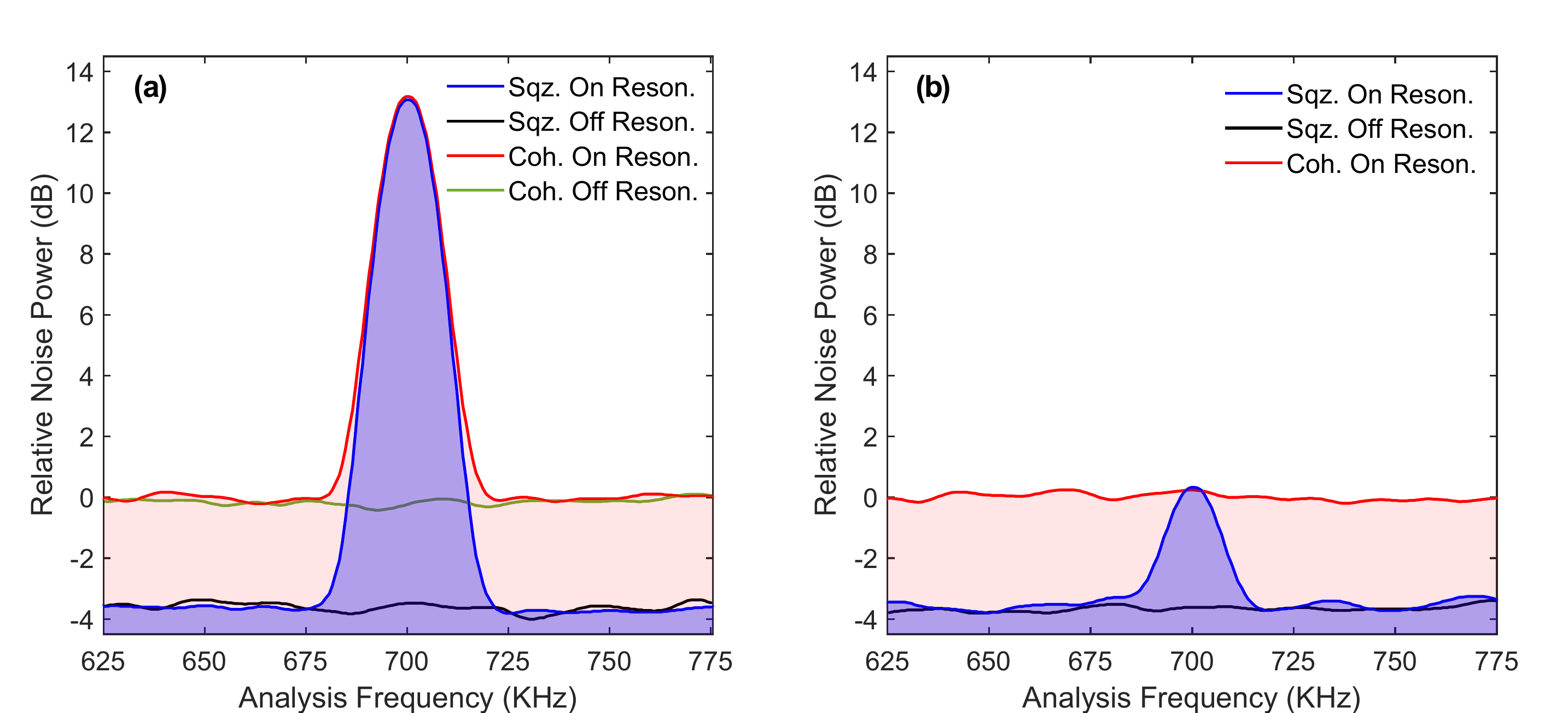}
    \caption{Demonstration of quantum-enhanced SBS spectroscopy using probe power of 750~$\mu$W and pump power of (a) 36~mW and (b) 7.5~mW. The red and blue SBS traces correspond to the configurations where the probe beam is in a coherent state and in a two-mode squeezed state respectively. 
    %with -~3.40~dB noise reduction below the shot-noise level. 
    The green and black traces correspond to the realization where the two lasers are locked outside of the water SBS gain profile. All traces are normalized to the shot-noise level. 
        \label{2Pumps}}
    %\end{center}
\end{figure}

The quantum-enhanced water SBS spectra can also be acquired using a RF spectrum analyzer. The experimental results are presented in Fig.~\ref{2Pumps}. In order to acquire the spectra, both lasers are locked so that their frequency difference matches the Brillouin shift of water, which in our case is 5~GHz, indicated by the gain peak in Fig.~\ref{LockIn}(a). The data presented in Fig.~\ref{2Pumps} and the following graphs have all been measured by a RF spectrum analyzer with a resolution bandwidth of 10~KHz and a video bandwidth of 10~Hz. \textcolor{black}{With these bandwidths, the shot noise level indicated by the red curve is at -69.5~dBm, whereas the electronic noise floor is at -81~dBm, There is no contribution from the stray pump light to the detection noise}. We present the spectra for the Brillouin gain of water using coherent beams (red traces) and twin beams (blue traces) with 750~$\mu$W probe power, while pump power is kept at 36~mW in Fig.~\ref{2Pumps}(a) and 7.5~mW in Fig.~\ref{2Pumps}(b).  It is clear from the spectra that the implementation of twin beams significantly improves the SNR of the SBS gain, and therefore the sensitivity of the Brillouin spectroscopy. We see in particular in Fig.~\ref{2Pumps}(b) that for pump power of 7.5~mW, the Brillouin gain from two coherent beams is almost embedded in shot-noise and only becomes pronounced when using twin beams. It is therefore clear that by using the two-mode squeezed light, it is possible to obtain Brillouin gain even for CW pump powers less than 8~mW. This is extremely beneficial when studying fragile biological samples where excessive optical power might damage the sample.

\begin{figure}[t]
    %\begin{center}
    \includegraphics[width=1\linewidth]{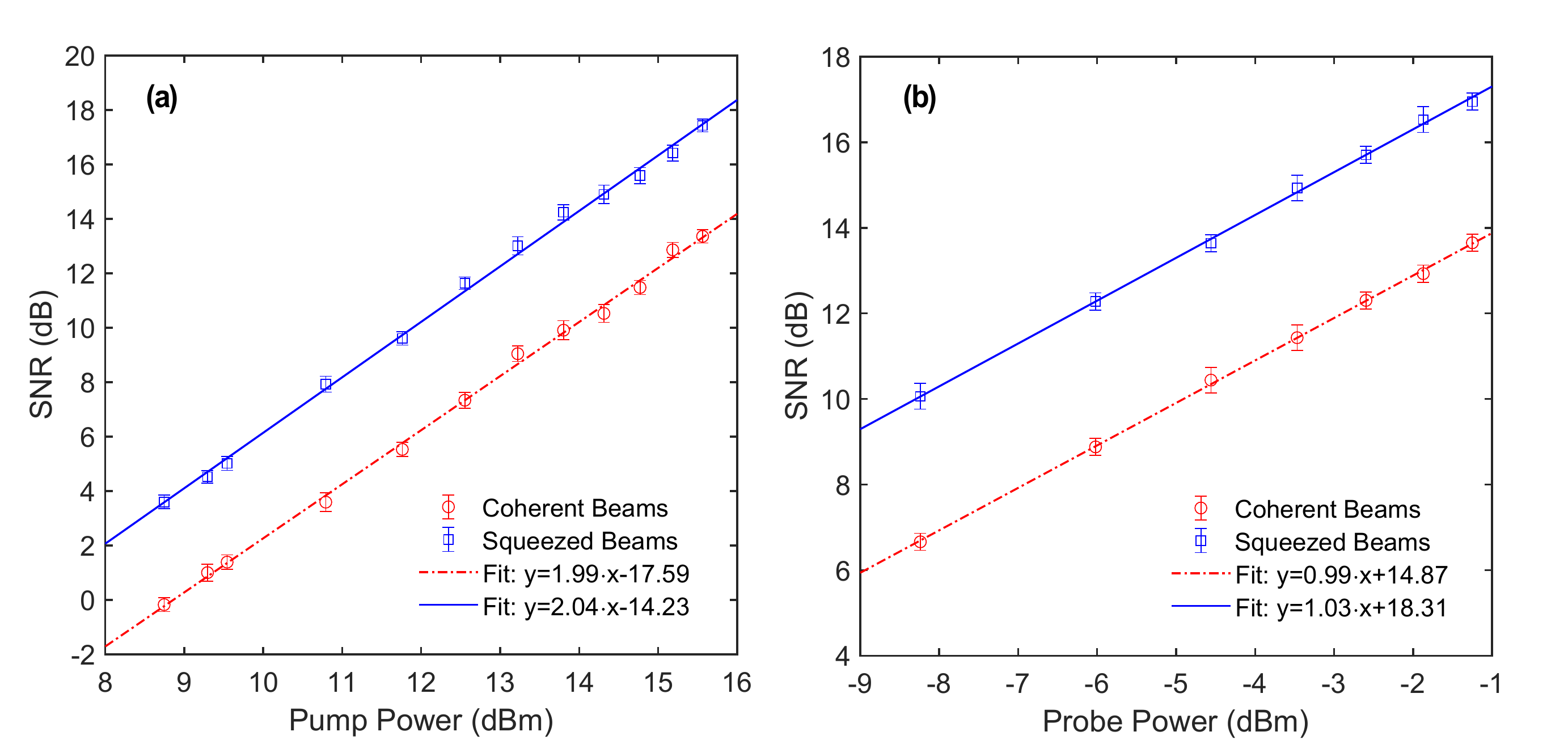}
    \caption{SNR of the SBS gain peak of water as a function of optical power (in dBm) of (a) pump beam \textcolor{black}{and (b) probe beam}. The red circles and dot-dashed fit line correspond to the probe beam being in a coherent state, while the blue squares and solid fit line correspond to the probe beam being in a two-mode squeezed state. 
             %The average quantum advantage can be calculated from the fitting parameters as $17.59-14.23 = 3.36$~dB. 
    \label{SNR}}
    %\end{center}
\end{figure}

We also plot in Fig.~\ref{SNR} the SNR of the water SBS gain as a function of optical power (in dBm) of \textcolor{black}{the two input beams (Figs.~\ref{SNR}(a) and (b) are for the pump and probe beams respectively)} for the cases where the probe beam is in a coherent state (red circles) and in a two-mode squeezed state (blue squares). The probe beam power is kept at 750~$\mu$W in Fig.~\ref{SNR}(a), \textcolor{black}{and the pump beam power is kept at 36~mW in Fig.~\ref{SNR}(b).} The error bars correspond to one standard deviation. From the fits we see nice linear dependence of the SBS gain on the pump power with slope of 1.99 and 2.04 in Fig.~\ref{SNR}(a), \textcolor{black}{and on the probe power with slope of 0.99 and 1.03 in Fig.~\ref{SNR}(b), which matches our expectations of 2 and 1 for Figs.~\ref{SNR}(a) and (b) respectively~\cite{SM}.} Also notice that, the average noise suppression (in dB) below the shot-noise level can be calculated from the fitting parameters as $17.59-14.23 = 3.36$~dB \textcolor{black}{and $18.31-14.87=3.44$~dB for Figs.~\ref{SNR}(a) and (b) respectively.} This is the quantum advantage of the two-mode squeezed light over coherent light in the SNR of gain measurement for the SBS spectroscopy. \textcolor{black}{Note that in terms of the \textit{sensitivity} of the gain measurement, our scheme is also \textit {sub-shot-noise-limited}, which is $\sim 1.7$~dB below SNL~\cite{PhysRevApplied.15.044030,SM}.} However, for applications in biological imaging, the incident intensity is always limited by a number of effects related to phototoxicity, thus even a tiny enhancement ($< 1$~dB), as it was recently shown by Casacio \textit{et al.}~\cite{casacio2021quantum}, makes a huge difference in improving image contrast, which has a tremendous impact in biological imaging.

\begin{figure}[t]
    %\begin{center}
    \includegraphics[width=1\linewidth]{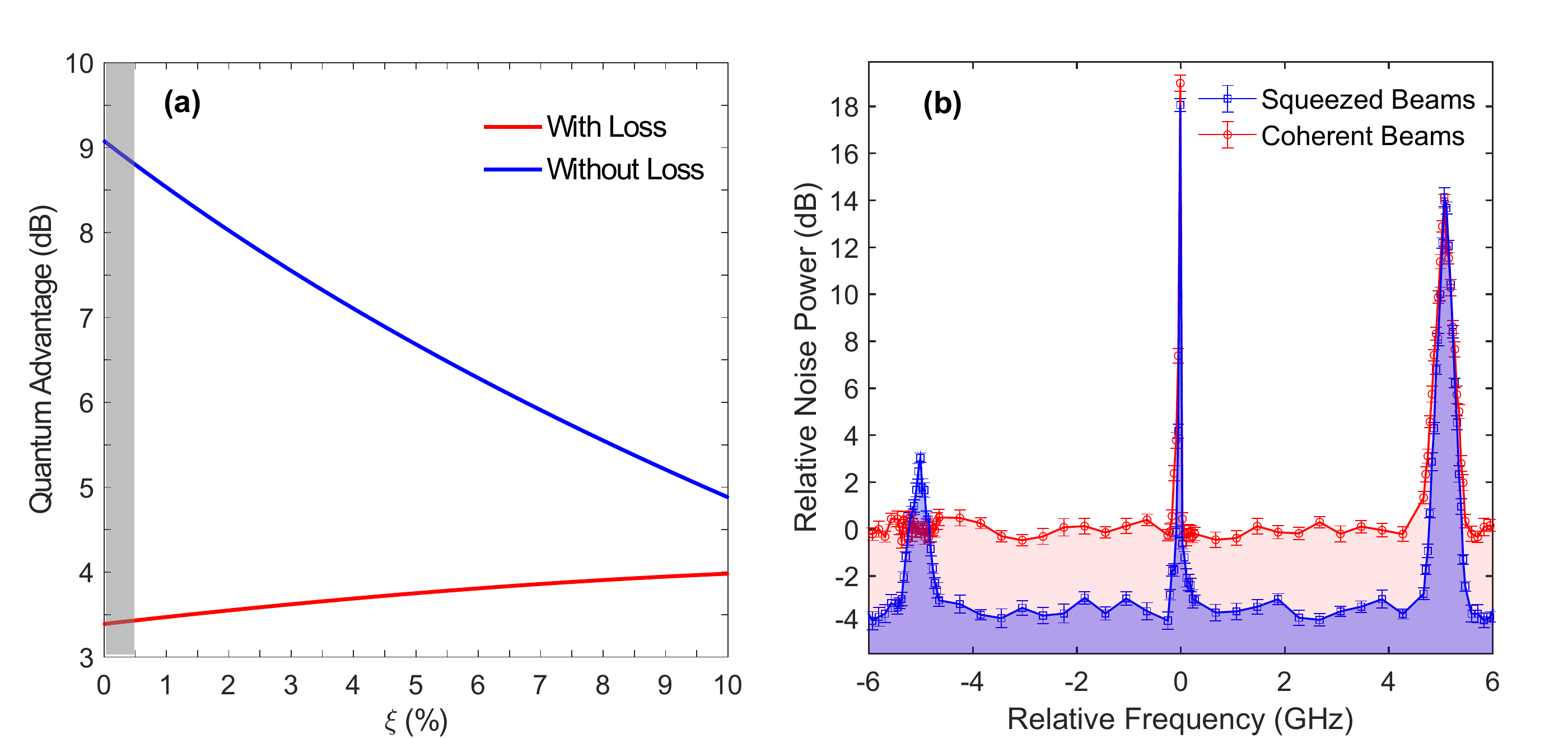}
    \caption{(a) Theoretical prediction for the quantum advantage as a function of the SBS gain related parameter $\xi=\text{G}_{\text{}}-1$. The red curve is plotted with experimental conditions, while the blue curve is plotted with no loss present. (b) SBS spectrum of water acquired using a RF spectrum analyser. 
    %by scanning the locking frequency of the pump laser. 
    %All other experimental conditions are the same as in Fig.~\ref{2Pumps}(a).
    \label{Spectrum}}
    %\end{center}
\end{figure}

This $\sim$~3.40~dB quantum advantage can be calculated with a theoretical framework assuming that both twin beams are in single modes~\cite{SM}. 
The theoretical quantum advantage (i.e., the improvement in SNR) as a function of $\xi=\text{G}_{\text{}}-1$, where G is the SBS gain, is shown in Fig.~\ref{Spectrum}(a) as the red curve. Since $\xi$ is in the range of $10^{-6}$ to $10^{-5}$~\cite{SM} (within the region highlighted by the gray bar), thus the measured $\sim$~3.40~dB quantum advantage agrees very well with our theoretical prediction. 
%As a comparison, we also plot the theoretical curve without loss 
%($\eta_{\text{p1}}=\eta_{\text{p2}}=\eta_{\text{c}}=0$) 
%in blue in Fig.~\ref{Spectrum}(a) to show the effect of the SBS gain on the quantum advantage. 
%We see from these two curves that, 
Also note that with loss absent, as shown by the blue curve, the SBS gain only degrades the quantum advantage because the gain process itself introduces noise, whereas with loss present, the SBS gain instead improves the degradation of quantum advantage due to the competition between the gain and loss.

The SBS spectrum of water acquired by a lock-in amplifier shown in Fig.~\ref{LockIn}(a) can also be attained using the RF spectrum analyser. The results are shown in Fig.~\ref{Spectrum}(b). This requires scanning the locking frequency of the pump laser while keeping the locking frequency of the probe laser fixed. 
%The results are plotted in Fig.~\ref{Spectrum} with red circles being the probe in a coherent state and blue squares being the probe in the two-mode squeezed state. 
All data points are obtained by normalizing the water SBS gain at 700~KHz to the shot-noise level, and error bars correspond to one standard deviation. The powers of the two laser beam are the same as in Fig.~\ref{2Pumps}(a). A quantum advantage of $\sim$~3.40~dB can be clearly seen from these two spectra. It is also worth pointing out that the classical approach is not able to detect the SBS loss dip at $-5$~GHz \textcolor{black}{using a spectrum analyzer} as small differential absorption of two coherent beams would always be at the shot-noise level. \textcolor{black}{Whereas the SBS loss is observable with a lock-in amplifier, as shown in Fig.~\ref{LockIn}(a).}

\begin{figure}[t]
    %\begin{center}
    \includegraphics[width=1.03\linewidth]{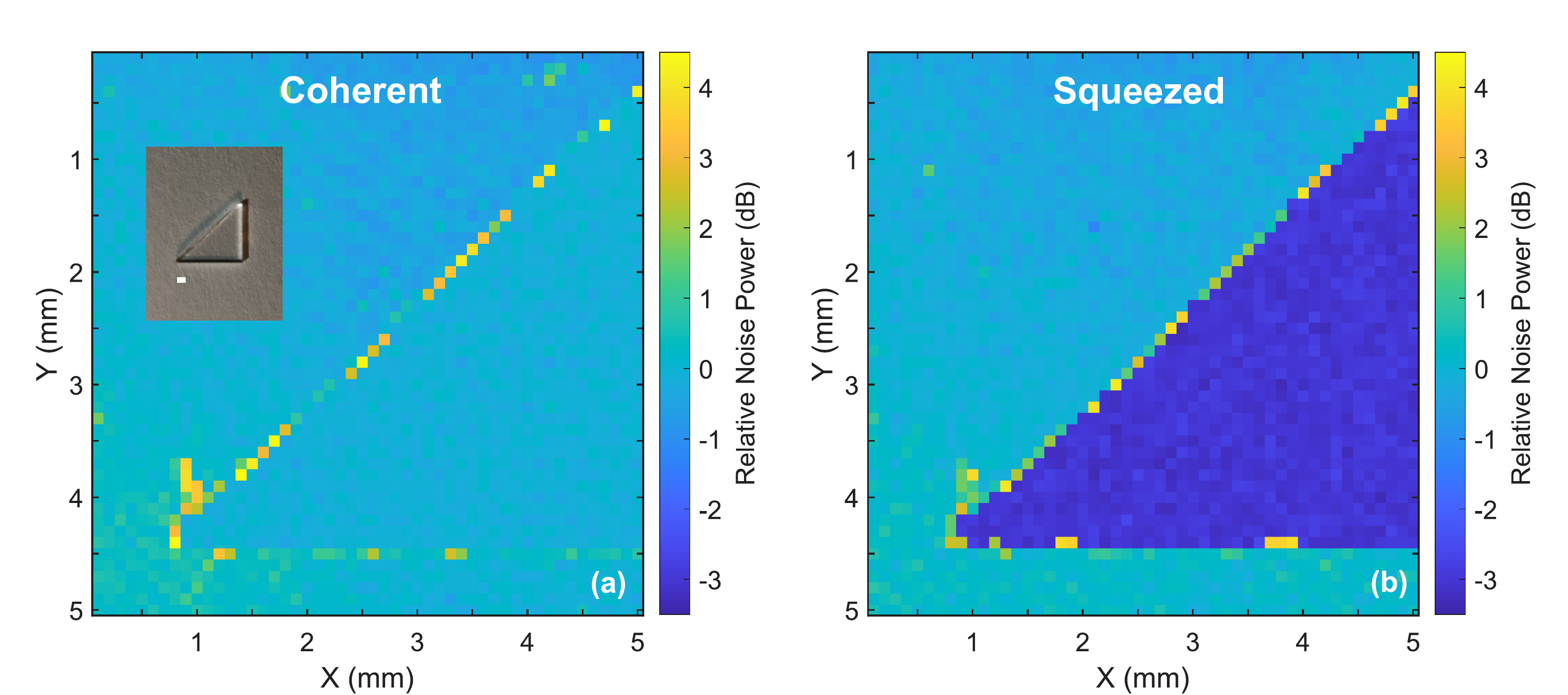}
    \caption{Quantum enhanced microscopic imaging using water as the signal medium. The imaging object is a triangle-shaped piece of glass shown in the inset of (a) where the white scale bar is 1 mm in horizontal direction. 
    %See text for a detailed description.
    %The images are acquired by electronically moving the sample using two translational stages with a special step size of 100~$\mu$m in both directions. All pixels are taken under the experimental conditions shown in Fig.~\ref{2Pumps}(b). Namely, pixels in (a) are registered with the probe beam being in a coherent state, which corresponds to the red curve in Fig.~\ref{2Pumps}(b), while pixels in (b) are registered with the probe beam being in the two-mode squeezed state, which corresponds to the blue curve in Fig.~\ref{2Pumps}(b). Each pixel takes about 2~s to obtain, which is limited by the sweep time of the RF spectrum analyzer. 
    More than 3~dB quantum-enhanced SNR, or image contrast, is clearly visible in (b). 
    %\end{center}
    \label{Image}}
\end{figure}

As the final step, we demonstrate that our scheme can also be utilized for microscopic imaging. We use the SBS gain of water to acquire a 2-Dimensional image of a piece of triangle shaped glass, as shown in the inset of Fig.~\ref{Image}(a). The pump and probe powers are 7.5~mW and 750~$\mu$W respectively (further image acquisition details can be found in Ref.~\cite{SM}). 
%The glass was made by cutting off a corner of a microscope slide whose thickness is 1~mm, and the lengths of the triangle's two sides are 7.5~mm and 6.5~mm respectively. Since our homemade sample holder consists of two glass microscope slides separated by 1~mm with water filled between them, the glass triangle therefore can be introduced into the sample holder so that there is no water content within the area of the triangle. 
%We use two translational stages with differential micrometer screws to automatically move the sample holder's position with a spatial scan step size of 100~$\mu$m in both directions. The images in Fig.~\ref{Image} are acquired by scanning each pixel under the experimental conditions shown in Fig.~\ref{2Pumps}(b). Namely, the pump and probe powers are 7.5~mW and 750~$\mu$W respectively, and the two lasers are locked so that their frequency difference matches the 5~GHz Brillouin shift of water.
Pixels in Fig.~\ref{Image}(a) and Fig.~\ref{Image}(b) are registered with the probe beam being in a coherent state and the two-mode squeezed state respectively. Obviously, the image contrast (i.e., the SNR) for the glass triangle in Fig.~\ref{Image}(a) is unappreciable due to the coherent light induced SBS gain of water is overwhelmed by the shot noise (see the red curve in Fig.~\ref{2Pumps}(b)). By using the two-mode squeezed light, however, a clear image contrast of more than 3~dB for the glass triangle is obtained in Fig.~\ref{Image}(b) (see the blue curve in Fig.~\ref{2Pumps}(b)). \textcolor{black}{Notice that each pixel in Fig.~\ref{Image} takes $\sim$~2~s to obtain, this seemly slow acquisition time is \textit {not fundamentally limited by our scheme itself}, but is rather technically limited by the instrument (i.e., the ‘write’ and ‘read’ time of the instrument-computer interface). In principle, our acquisition rate can be readily improved to be 240~ms per 4~GHz spectrum, with only $8.25$~mW total excitation power \textcolor{black}{(see Ref.~\cite{remer2020high} for the state-of-the-art SBS imaging performance, and Ref.~\cite{SM} for a detailed discussion on our scheme's usefulness and limitations).}}
%which is mainly limited by the ‘write’ and ‘read’ time of the instrument-computer interface, i.e., one needs to wait for at least 2 s for the spectrum analyzer to write the trace to its memory before the computer can view/read from it. This implies that, our seemly slow acquisition rate is not fundamentally limited by our scheme itself, but is rather technically limited by the instrument. This technical limitation can be readily circumvented by the use of a more advanced RF spectrum analyzer (i.e., a real-time spectrum analyzer) having a much faster data writing and read-out rate, or having a large memory that data can be stored and processed \textit{locally}, so that the acquisition time would be solely limited by the sweep time of the spectrum analyzer. In principle, our acquisition rate can be readily improved to be 240~ms per 4~GHz spectrum with only $8.25$~mW infrared excitation power (see Ref.~\cite{SM} for a detailed discussion).}

\textcolor{black}{Note that our sample is composed of only two components - glass and water. Since glass has no Brillouin peak around 5~GHz, the SBS signal only comes from water by scanning the laser around 5~GHz, therefore, if we were to acquire the image of a glass triangle by presenting the Brillouin shift estimation, that would look almost the same as Fig.~\ref{Image}, only difference is that the acquisition time would be significantly extended due to the need to acquire the entire spectrum at each pixel. In this work, we follow the earlier established protocol for SBS microscopy (Ref.~\cite{ballmann2015stimulated}) to display the intensity of signal taken at the peak of Brillouin spectrum for a given material as a function of position.}

\textcolor{black}{It is also worth mentioning that, our probe power can go up to almost 5~mW. The 750~$\mu$W upper bound is chosen here simply because the balanced detector would be saturated at 1~mW of input power. This limitation is, again, not fundamental to our scheme, an \textit{AC-coupled} balanced detector would overcome it. The lateral and axial resolutions in this work are 5~$\mu$m and 70~$\mu$m respectively, which can also be readily improved to be 1.5~$\mu$m and 4.5~$\mu$m respectively using molded aspheric lenses (see Ref.~\cite{SM} for a detailed discussion).}

In conclusion, we demonstrated a quantum-enhanced continuous-wave SBS spectroscopy and imaging scheme. As a proof-of-principle, we acquired a SBS spectrum of water and a 2-dimensional microscopic image with quantum enhanced SNR/contrast of $\sim$~3.4~dB. %Our setup utilized two amplitude-modulated CW lasers, achieving as little as 7.5~mW and 750~$\mu$W average power on the sample from the pump and probe beams respectively. 
The quantum enhancement is achieved by using the two-mode intensity-difference squeezed light with a spectral width in the range of 10~MHz generated by the FWM process in atomic $^{85}$Rb vapor. It is very important to note that, it is this unique narrow-band feature of our squeezed light that makes the quantum-enhanced SBS spectroscopy and imaging system possible, as for the SBS process to ocurr the spectral width of the light source must be well below the Brillouin linewidth ($\sim 300$~MHz in this work). The low optical power (can be $<$~8~mW) and the excitation wavelengths in the water transparency window used in this work has made our system very applicable for probing mechanical properties of biological samples, which will be the subject of our future study.

% We gratefully acknowledge the support of Air Force Office of Scientific Research (Awards No. FA-9550-20-1-0366, FA9550-15-1-0517, and FA9550-20-1-0367), the National Science Foundation (NSF) (Awards No. DBI-1455671, ECCS-1509268 and CMMI-1826078), Army Medical Research (Grant No. W81XWH2010777), the National Institutes of Health (NIH) (Grant No. 1R01GM127696 and 1R21GM142107), the Cancer Prevention and Research Institute of Texas (CPRIT) (Award No. RP180588), and the Robert A. Welch Foundation (Grant No. A-1943). 

{\bf Funding.} Air Force Office of Scientific Research (Awards No. FA-9550-20-1-0366 and FA-9550-20-1-0367), the National Science Foundation (NSF) (Awards No. CMMI-1826078), Army Medical Research (Grant No. W81XWH2010777), the National Institutes of Health (NIH) (Grant No. 1R01GM127696, 1R21GM142107, and 1R21CA269099), the Cancer Prevention and Research Institute of Texas (CPRIT) (Award No. RP180588), and the Robert A. Welch Foundation (Grant No. A-1943).

{\bf Disclosures.} The authors declare no conflicts of interest.

{\bf Data availability.} Data underlying the results may be obtained from the authors upon reasonable request.

{\bf Supplemental document.}
See Supplement Material for supporting content.

\section{Experimental Details}

\subsection{Optical Arrangement Description}

\begin{figure}[]
    \begin{center}
    \includegraphics[width=1\linewidth]{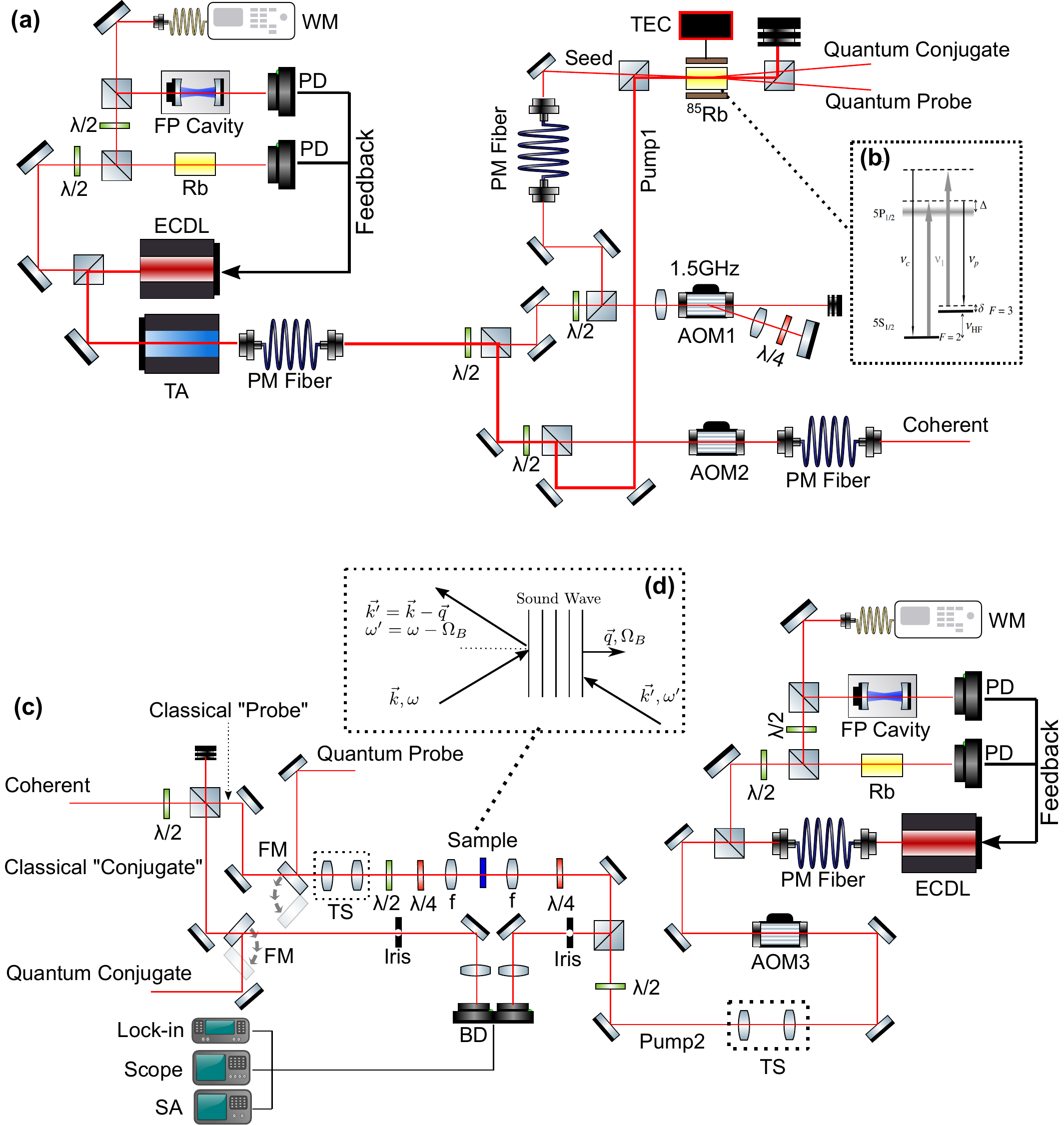}
    \caption{\textcolor{black}{
     (a) Experimental setup for the bright two-mode squeezed light generation based on FWM in $^{85}$Rb atomic vapor. See text for a detailed description.
     %in which a seeded $^{85}$Rb vapor cell produces strong quantum-correlated twin beams (``Probe" and ``Conj.") via FWM. The twin beams are separated from the pump (``Pump 1") by a $\sim$~$2\times10^5$~$:1$ Glan-Laser polarizer after the cell. The probe beam and another pump beam (``Pump 2") form a geometry for the SBS process in water (see text for details) while the conjugate beam serves as a reference, before they are focused onto a balanced detector (BD). Two flip mirrors (FM) are used to steer in two coherent beams to form a classical version of the whole setup. The three AOMs in the beam paths are used to amplitude-modulate individual beams. 
     ECDL: external cavity diode laser, TA: tapered amplifier, FP cavity: Fabry-Perot cavity, FM: flip mirror, TS: telescope, BD: balanced detector, PD: photodiode, AOM: acousto-optic modulator; PM fiber: polarization maintaining fiber, TEC: thermo-electric cooler, WM: wave meter, SA: RF spectrum analyser. (b) Level structure of the D1 transition of $^{85}$Rb atom. The optical transitions are arranged in a double~-~$\Lambda$ configuration, where $\nu_p$, $\nu_c$ and $\nu_1$ stand for probe, conjugate and pump frequencies, respectively, fulfilling $\nu_p$ +  $\nu_c$ =  $2\nu_1$ and $\nu_c - \nu_p = 2\nu_{HF}$. The width of the excited state in the level diagram represents the Doppler broadened line. $\Delta$ is the one-photon detuning. $\nu_{\text{HF}}$ is the hyperfine splitting in the electronic ground state of $^{85}$Rb. (c) Experimental setup for the SBS spectroscopy for both the quantum and classical configurations. (d) Phase-matching diagram for the SBS process~\cite{NonlinearOptics}. The wave-vectors and frequencies for the pump, probe and sound wave are denoted by ($\vec{k}$, $\omega$), ($\vec{k^{\prime}}$, $\omega^{\prime}$) and ($\vec{q}$, $\Omega_B$) respectively.}
        \label{Setup}}
    \end{center}
\end{figure}

The atomic medium is pumped by a strong ($\sim 500$~mW) narrow-band continuous-wave (CW) laser \textcolor{black}{composed of an external cavity diode laser (ECDL) and a tapered amplifier (TA)} (shown in Fig.~\ref{Setup}(a) as ``Pump 1") at frequency $\nu_1$ ($\lambda = 795$~nm) with a typical linewidth $\Delta \nu_1 < 1$~MHz. Applying an additional weak (in the range of a few hundreds $\mu$W) coherent beam (shown in Fig.~\ref{Setup}(a) as ``Seed") %of $\sim 1$~mW optical power, 
at frequency $\nu_p = \nu_1 - (\nu_{HF}+\delta)$, where $\nu_{HF}=3.036$~GHz and $\delta$ are the hyperfine splitting in the electronic ground state of $^{85}$Rb and the two-photon detuning ($\delta=5$~MHz in this work) respectively in Fig.~\ref{Setup}(b) . \textcolor{black}{The frequency difference between ``Pump 1'' and ``Seed'' is acquired by double-passing an 1.5~GHz acousto-optic modulator (AOM) (shown in Fig.~\ref{Setup}(a) as ``AOM1").} Two pump photons are converted into a pair of twin photons, namely `probe $\nu_p$' and `conjugate $\nu_c$' photons, adhering to the energy conservation $2 \nu_1 = \nu_p + \nu_c$ (see the level structure in Fig.~\ref{Setup}(b)). The resulting twin beams are strongly quantum-correlated and are also referred to as \textit{bright two-mode squeezed light}~\cite{PhysRevA.78.043816}. The twin beams exhibit a intensity-difference squeezing of 6.5~dB measured by a balanced detector \textcolor{black}{(with customized photodiodes having 94~\% quantum efficiency at 795~nm,} further squeezing measurement details can be found in Ref.~\cite{PhysRevApplied.15.044030}), which is indicative of strong quantum correlations~\cite{PhysRevA.78.043816}.\\

\textcolor{black}{The Pump1 and
Seed beams are combined in a polarising beam splitter and directed at an angle of $\sim0.3^\circ$ to each other into a 12.5~mm long vapor cell filled with isotopically pure $^{85}$Rb. The two beams are collimated with 700~$\mu$m and 400~$\mu$m $1/e^2$ waists at the cell center respectively. The cell, with no magnetic shielding, is kept at $105^\circ$C by a thermo-electric cooler (TEC) and a PID feedback loop. The windows of the cell are anti-reflection coated on both faces, resulting in a transmission for the Seed beam of $\sim98$\% per window.} After the $^{85}$Rb vapor cell,  the Pump1 and the twin beams (shown in Fig.~\ref{Setup}(a) as ``Probe" and ``Conj.") are separated by a Glan-Laser polarizer, with $\sim$~$2\times10^5$~$:1$ extinction ratio for the Pump1. The probe beam then passes through a telescope (TS) with an enlarged beam waist ($\sim 3$~mm) before focused (down to a $1/e^2$ beam waist of $\sim 5$~$\mu$m by a plano-convex lens with focal length $f= 16$~mm) and overlapped with a counter propagating laser beam (shown in Fig.~\ref{Setup}(c) as ``Pump 2", a same type of ECDL as ``Pump 1", and having a $1/e^2$ beam waist of $\sim 6$~$\mu$m) at a homemade sample holder filled with distilled water, to form a phase-matching geometry for the SBS process in water depicted in Fig.~\ref{Setup}(d). The sample holder consists of two glass microscope slides separated by 1 mm with water filled between them. Both $\lambda/2$ and $\lambda/4$ waveplates are added in the probe beam path in order for the probe beam to be reflected as much as possible by the polarizing beam splitter (PBS) into one port of the balanced detector (BD). Therefore in this configuration, the probe beam is linearly polarized while the pump beam for the SBS process (``Pump 2" in Fig.~\ref{Setup}(c)) is circularly polarized. The conjugate beam serves as a reference, and two flip mirrors (FM) are used for the introduction of two coherent beams so that the whole setup can be converted into a classical version. The pumps and probe beams are amplitude-modulated by three AOMs at 300~KHz (AOM1) and 400~KHz (AOM2\&3) respectively. The water SBS signal therefore is expected to appear at 700~KHz where the two-mode squeezing is expected to be the best~\cite{PhysRevApplied.15.044030,doi:10.1063/5.0010909}. \textcolor{black}{The balanced detection uses two coherent beams and a balanced detector, which subtracts away common-mode noise to better than 25 dB, therefore contributions from low-frequency technical noise can be eliminated, so that the noise level at the modulation frequency (where the signal occurs) would be (or close to) shot-noise limited. There is no contribution from the stray pump light to the detection noise as there is a decent angle separation ($\sim0.3^\circ$) between the twin beams and the Pump1 beam, and there are multiple irises in the paths of the twin beams to filter out the stray pump light.}\\

\subsection{\textcolor{black}{Locking of the Two Lasers}}

\textcolor{black}{In addition to the components shown in Fig.~\ref{Setup}(a), there are also frequency-locking optics and electronics for the probe and pump beams of the SBS process so that they can be locked and separated by the phonon frequency $\Omega_B/2\pi$ (i.e., ``Brillouin frequency shift'') of water, which is in the range of $\sim 5$~GHz~\cite{ballmann2015stimulated}. In this work, we use fringes from a room temperature Fabry-Perot cavity as the locking error signal and absorption lines from a room temperature natural abundant Rb cell as the locking reference for each laser beam. Note that in Figs.~\ref{Setup}(a) and~\ref{Setup}(c), there are two Rb cells for locking, in reality they are the same one. We change the frequency difference between the two beams by fixing the probe frequency (blue-tuned by a ‘one-photon detuning $\Delta$’ of 900~MHz with respect to the $^{85}$Rb $5\text{S}_{1/2}, \text{F} = 2 \rightarrow 5\text{P}_{1/2}$, D1 transition shown in Fig.~\ref{Setup}(b)) so that the FWM process can yield the best two-mode squeezing, \textcolor{black}{while scanning the locking frequency of the pump with a minimal step of 40~MHz (determined by the resolution of the scanning voltage).}}\\

\subsection{\textcolor{black}{The Necessity of Two-Beam Modulation and Two-Beam Balanced Detection}}

\textcolor{black}{Note that in principle only one modulation on the pump beam would be sufficient for attaining stimulated Brillouin scattering signal. However, one must be extremely careful to eradicate any contribution from the pump light to the detected signal, as any amount of residual pump leakage into the detector would appear as spurious SBS signal. To do this, Ref.~\cite{remer2020high} used a Rubidium-85 notch filter at the pump frequency. This approach, however, would not be practical in our scheme as our two-mode squeezed twin beams, i.e., the probe and conjugate beams, are only a few GHz separated from the pump frequency (the pump here is for the SBS process, not the pump for the FWM process), therefore the use of any notch filter at the pump frequency would inevitably induce undesired atomic absorption at the probe and conjugate frequencies as well. This would significantly deteriorate the quantum correlations between the probe and conjugate beams, and eventually wear out the quantum advantage. The 2-beam modulation (pump and probe at 300~KHz and 400~KHz respectively) approach adopted in our scheme solved this issue, as the SBS signal appeared at the sum frequency 700~KHz, hence even if there is residual pump leakage into the detector, the ``spurious signal'' would only appear at 300~KHz.}\\

\textcolor{black}{It is very important to point out that in our scheme, we used a ``two-mode" squeezed state, where squeezing resides in the ``intensity-difference" between the two involving modes. As opposed to the experimental complexity of a single-mode squeezed scheme where a homodyne measurement is needed to characterize the squeezing, and a phase-locking mechanism is needed to track the squeezed quadrature, our scheme only requires a balance detector so that an intensity-difference measurement can be obtained. Therefore a ``balanced coherent detection" where two coherent beams are used would be the appropriate classical counterpart to the quantum configuration in our scheme.}\\

\subsection{Microscopic Imaging Acquisition}

The triangle-shaped glass used for imaging shown in the inset of Fig.~6(a) was made by cutting off a corner of a microscope slide whose thickness is 1~mm, and the lengths of the triangle's two sides are 7.5~mm and 6.5~mm respectively. Since our homemade sample holder consists of two glass microscope slides separated by 1~mm with water filled between them, the glass triangle therefore can be introduced into the sample holder so that there is no water content within the area of the triangle.

To acquire the images in Fig.~6 in the main text, we use two translational stages with differential micrometer screws to automatically move the sample holder's position with a spatial scan step size of 100~$\mu$m in both directions. The images are obtained by scanning each pixel under the experimental conditions shown in Fig.~3(b). Namely, the pump and probe powers are 7.5~mW and 750~$\mu$W respectively, and the two lasers are locked so that their frequency difference matches the 5~GHz Brillouin shift of water.

\section{Theoretical Framework}

We use a single-mode quantum-mechanical model to simulate the experiment~\cite{PhysRevResearch.3.033095}. 
%Similar treatments have already been implemented in Refs.~\cite{PhysRevResearch.3.033095,PhysRevApplied.15.044030,Li:17}. 
We denote the optical field operators for the probe and conjugate modes as $\hat{a}_{0,f}$ and $\hat{b}_{0,f}$ with subscripts $0$ and $f$ labeling the operators at the initial and final stages of transformation, respectively. The input-output relation for the FWM process, $\hat{a}_{\text{FWM}} = (\text{cosh} r)\hat{a} + (\text{sinh} r)\hat{b}^{\dagger}$, where $r$ is the squeezing operator, is well known. For the input-output relation for the SBS process, we can write it as $\hat{a}_{\text{SBS}} = g\hat{a}_{\text{FWM}} + \hat{\digamma}$, where $g$ is the SBS gain parameter, and $\hat{\digamma}$ is the noise operator introduced by the SBS gain process. The field operator $\hat{a}_{\text{SBS}}$ must satisfy the commutation relation $[\hat{a}_{\text{SBS}}, \hat{a}^{\dagger}_{\text{SBS}}]=1$, from which the noise operator $\hat{\digamma}$ can be derived as $\sqrt{g^2-1}\hat{\nu}_{\text{B}}^{\dagger}$, where $\hat{\nu}_{\text{B}}$ is a vacuum noise operator introduced by the SBS process. All optical and atomic absorption losses sustained by the twin beams are modeled by three beams splitters with transmission $\eta_{\text{p1}}$, $\eta_{\text{p2}}$ and $\eta_{\text{c}}$~\cite{Li:17}. They represent the atomic and optical loss in the probe pathway between the FWM cell and the SBS sample holder ($\eta_{\text{p1}}$), between the SBS sample holder and the balanced detector ($\eta_{\text{p2}}$), and the optical loss in the conjugate pathway ($\eta_{\text{c}}$), respectively. 
The experimentally measured values are  $r=1.39$ (which corresponds to our measured FWM gain of $\text{cosh}^2 r=4.5$), $\eta_{\text{p1}}=0.83$, $\eta_{\text{p2}}=0.75$ and $\eta_{\text{c}}=0.9$. \textcolor{black}{Since we have kept all the noise operators, thus it is not the case of `noiseless' amplification. Therefore this theoretical framework is not under the unitary condition.} We treat all pump beams classically. The vectors $\hat{\vec{{V}_0}}$ and $\hat{\vec{{V}_f}}$ are the initial and final field operators defined by

\begin{equation}
\hat{\vec{{V}_0}}=
\begin{pmatrix}
\hat{a}_0\\ \hat{a}^\dagger_0\\ \hat{b}_0\\ \hat{b}^\dagger_0
\end{pmatrix} \hspace{4mm} \text{and} \indent 
\hat{\vec{{V}_f}}=
\begin{pmatrix}
\hat{a}_f\\ \hat{a}^\dagger_f\\ \hat{b}_f\\ \hat{b}^\dagger_f
\end{pmatrix}.
\end{equation}
The experiment can then be described by the transformation of field operators 
\begin{equation}
\hat{\vec{{V}_f}}=\bm{T}_{\bm{2}} \cdot \left( \bm{B} \cdot \left[ \bm{T}_{\bm{1}} \cdot (\bm{F} \cdot  \hat{\vec{{V}_0}}) + \hat{\vec{{L}_1}}\right] + \hat{\vec{{L}_B}}\right) +\hat{\vec{{L}_2}},
\label{eq:physics}
\end{equation}
where 
\begin{equation}
%\begin{split}
\bm{F}=
\begin{pmatrix}
\cosh{r} &0 &0 &\sinh{r}\\ 0 &\cosh{r} &\sinh{r} &0\\ 0 &\sinh{r} &\cosh{r} &0\\ \sinh{r} &0 &0 &\cosh{r}
\end{pmatrix},
\end{equation}

\begin{equation}
\bm{B}=
\begin{pmatrix}
g &0 &0 &0\\ 0 &g &0 &0\\ 0 &0 &1 &0\\ 0 &0 &0 &1 
\end{pmatrix}, 
%\end{split}
\end{equation}
and 
\begin{equation}
\hat{\vec{{L}_B}}=
\begin{pmatrix}
\sqrt{g^2-1}\hat{\nu}_{\text{B}}^{\dagger}\\ \sqrt{g^2-1}\hat{\nu}_{\text{B}}\\ 0 \\ 0 
\end{pmatrix}.
%\end{split}
\end{equation}
% respectively. Here, $r$ and $s$ are the squeezing parameters which are related to the gains of the source and PSA via $G_{\text{source}} = \cosh^2 r$ and $G_{\text{PSA}} = e^{2 s}$. $\phi_{\text{PSA}}$ is the phase of the PSA defined in Eq.~(\ref{eq:phase}). We keep $\phi_{\text{PSA}} = 0$ such that the intensity gain of the probe beam is always $G_{\text{PSA}}$.
Matrix $\bm{F}$ describes the FWM process, while matrix $\bm{B}$ together with vector $\hat{\vec{{L}_B}}$ describe the SBS process. Matrices $\bm{T}_{\bm{1}}$ and $\bm{T}_{\bm{2}}$ describe the transmission of the beam splitters, and vectors $\hat{\vec{{L}_1}}$ and $\hat{\vec{{L}_2}}$ contain the field operators $\hat{\mu}_{\text{p1}}$, $\hat{\mu}_{\text{p2}}$ and $\hat{\mu}_{\text{c}}$ for the vacuum noise coupled in by optical losses:

\begin{equation}
\bm{T}_{\bm{1}}=
\begin{pmatrix}
\sqrt{\eta_{\text{p1}}}& 0& 0& 0\\ 0& \sqrt{\eta_{\text{p1}}}& 0& 0\\ 0& 0& 1& 0 \\ 0& 0& 0& 1 
\end{pmatrix},
\end{equation}

\begin{equation}
%\begin{split}
\bm{T}_{\bm{2}}=
\begin{pmatrix}
\sqrt{\eta_{\text{p2}}}& 0& 0& 0\\ 0& \sqrt{\eta_{\text{p2}}}& 0& 0\\ 0& 0& \sqrt{\eta_{\text{c}}}& 0 \\ 0& 0& 0& \sqrt{\eta_{\text{c}}} 
\end{pmatrix}, 
\end{equation}

\begin{equation}
\hat{\vec{{L}_1}}=
\begin{pmatrix}
i\sqrt{1-\eta_{\text{p1}}}\hat{\mu}_{\text{p1}}\\ -i\sqrt{1-\eta_{\text{p1}}}\hat{\mu}^\dagger_{\text{p1}}\\ 0 \\ 0 
\end{pmatrix}.
%\end{split}
\end{equation}

\begin{equation}
\hat{\vec{{L}_2}}=
\begin{pmatrix}
i\sqrt{1-\eta_{\text{p2}}}\hat{\mu}_{\text{p2}}\\ -i\sqrt{1-\eta_{\text{p2}}}\hat{\mu}^\dagger_{\text{p2}} \\ i\sqrt{1-\eta_{\text{c}}}\hat{\mu}_{\text{c}}\\ -i\sqrt{1-\eta_{\text{c}}}\hat{\mu}^\dagger_{\text{c}}
\end{pmatrix}.
\end{equation}
When a coherent state $\vert\beta\rangle$, $\beta=\vert\beta\vert e^{i\phi}$, where $\phi$ is the input phase, seeds mode $a$, and only vacuum fluctuations $|0\rangle$ seed mode $b$, then the input state can be written as $\vert\beta, 0, 0, 0, 0, 0\rangle$, where the last four zeros are inputs for the vacuum/noise operators $\hat{\nu}_{\text{B}}$, $\hat{\mu}_{\text{p1}}$, $\hat{\mu}_{\text{p2}}$ and $\hat{\mu}_{\text{c}}$ respectively. Although not trivial, it is fairly straightforward to calculate the number operators $\hat {n}_a = \hat{a}^{\dagger}_f\hat{a}_f$ and $\hat{n}_b=\hat{b}^{\dagger}_f\hat{b}_f$ for the probe and conjugate modes after detection, and the expectation values of quantities such as the noise suppression below the shot noise level, i.e. the quantum advantage:

\begin{equation}
\text{Quantum Advantage [dB]} = -10 \times\text{log}_{10} \left[\frac{\Delta^2 (\hat{n}_a - \hat{n}_b)}{\Delta^2 \hat{n}_{\text{SNL}}} \right],
\label{eq_sq}
\end{equation}
where $\Delta^2 \hat{n}_{\text{SNL}}$ is the shot noise level, which is defined as the variance of the intensity difference of two coherent beams having the same intensities as the measured probe and conjugate beams, therefore in our case $\Delta^2 \hat{n}_{\text{SNL}} = \langle \hat{n}_a \rangle + \langle \hat{n}_b \rangle$. With the measured FWM gain and optical losses, theoretical curves shown in Fig.~4(b) in the main text can be thus readily plotted.

\section{SBS intensity gain parameter $\xi$ and signal magnitude estimation}

The SBS signal after interaction length $\text{L}$ can be written as 
\begin{equation}
\text{I}_{\text{out}}=\text{I}_{\text{probe}} \cdot e^{g_0 \text{I}_{\text{pump}} \text{L}},
\label{SBS1}
\end{equation}
where $g_0$ is the maximal SBS gain~\cite{ballmann2015stimulated}. In our experiment, $g_0 = 0.048~\text{m}/\text{GW}$~\cite{NonlinearOptics}, $\text{L}\cong 2 z_{\text{R}} \cong 70~\mu\text{m}$, where $z_{\text{R}}$ is the Rayleigh range as the signal comes almost entirely from the region where the intensity is the largest, and $\text{I}_{\text{pump}} \sim 1.2~ \text{GW}/\text{m}^2$ given the experimentally achievable maximal pump power of 36~mW and $1/e^2$ beam waist of $6~\mu$m at the focal point. This yields $g_0 \text{I}_{\text{pump}} \text{L} = 4.2 \times 10^{-6}$. Therefore 
\begin{equation}
\text{I}_{\text{out}} = \text{I}_{\text{probe}} \cdot e^{4.2 \times 10^{-6}} \cong (1+4.2 \times 10^{-6})\cdot\text{I}_{\text{probe}} = \text{G}_{\text{SBS}} \cdot \text{I}_{\text{probe}}.
\label{SBS2}
\end{equation}
Since we define in the main text that $\xi= \text{G}_{\text{SBS}} - 1$, thus $\xi = 4.2 \times 10^{-6}$, which is within the range indicated by the gray bar in Fig.~4(b) in the main text. \\

Since $\text{I}_{\text{out}}= (1+4.2 \times 10^{-6})\cdot\text{I}_{\text{probe}} = \text{I}_{\text{probe}} + 4.2 \times 10^{-6}\cdot\text{I}_{\text{probe}} = \text{I}_{\text{common mode}}+\text{I}_{\text{signal}}$, by using a balanced detector, the common mode intensity $\text{I}_{\text{probe}}$ is rejected, therefore we see that the maximum SBS signal going into the detector is $4.2 \times 10^{-6}$ of the total probe power (750~$\mu$W) going into the detector. The balanced detector has an electronic gain of $\sim 10^{5}$~V/W, therefore this SBS signal is $\sim 4.2 \times 10^{-6} \times 750 \times 10^5 = 315~\mu$V. As shown by the highest point in Fig.~2(b) in the main text, this estimation agrees with the measurement very well.

\section{Derivation of the relationships between SNR and the pump and probe optical powers}

In this section, we derive the relationship between the SNR (in dB) and the pump power (in dBm) of the SBS process. From Eq.~(\ref{SBS1}), we see that when $g_0\text{I}_{\text{pump}}\text{L} \ll 1$,
\begin{equation}
\text{I}_{\text{out}}=\text{I}_{\text{probe}} \cdot e^{g_0 \text{I}_{\text{pump}} \text{L}} \cong \text{I}_{\text{probe}} \cdot (g_0\text{I}_{\text{pump}}\text{L}) = \alpha \cdot \text{I}_{\text{probe}} \cdot \text{I}_{\text{pump}},
\label{SBS3}
\end{equation}
with constant $\alpha = g_0 \text{L}$. This relationship is demonstrated in Fig.~2(b) in the main text by keeping $\text{I}_{\text{pump}}$ unchanged. If we write noise on the SBS signal as 
\begin{equation}
\Delta \text{I}_{\text{out}}=\beta \cdot \sqrt{\text{I}_{\text{probe}}},
\label{SBS4}
\end{equation}
where $\beta$ is the noise factor, then for coherent excitation $\beta_{\text{coh}} = 1$, while for squeezed excitation $\beta_{\text{sqz}} < 1$. The SNR of this SBS signal registered by a RF spectrum analyser can thus be written as 
%\begin{equation}
\begin{align}
\text{SNR[dB]} &= 10 \times \text{log}_{10}(\frac{\text{I}_{\text{out}}}{\Delta \text{I}_{\text{out}}})^2 \\
&= 10 \times \text{log}_{10}(\frac{\alpha \cdot \text{I}_{\text{probe}} \cdot \text{I}_{\text{pump}}}{\beta \cdot \sqrt{\text{I}_{\text{probe}}}})^2\\
&= 20 \times \text{log}_{10}(\text{I}_{\text{pump}}) + 10 \times \text{log}_{10}(\frac{1}{\beta^2} \cdot \alpha^2 \text{I}_{\text{probe}})\\
&= 10 \times \text{log}_{10}(\text{I}_{\text{probe}}) + 10 \times \text{log}_{10}(\frac{1}{\beta^2} \cdot \alpha^2 \text{I}^2_{\text{pump}}).
\label{SBS5}
\end{align}
%\end{equation}
Therefore for a fixed probe power (i.e., $\alpha^2 \text{I}_{\text{probe}}$ is a constant), the SNR has a linear dependence on $10 \times \text{log}_{10}(\text{I}_{\text{pump}})$ (i.e., pump power in the unit of dBm) with a slope of 2. \textcolor{black}{On the other hand, for a fixed pump power (i.e., $\alpha^2 \text{I}^2_{\text{pump}}$ is a constant), the SNR has a linear dependence on $10 \times \text{log}_{10}(\text{I}_{\text{probe}})$ (i.e., probe power in the unit of dBm) with a slope of 1. These linear behaviors are demonstrated in Figs.~4(a) and~(b) respectively in the main text.} It is also worth noticing that information about the noise factor $\beta$ is contained in the second terms of Eqs.~(17) and~(\ref{SBS5}). In our case, the quantum advantage of using squeezed light for the SBS spectroscopy over coherent light would be simply the differences between the two second terms, 
\begin{align}
\text{Quantum Advantage [dB]} &= 10 \times \text{log}_{10}(\frac{1}{\beta_{\text{sqz}}^2} \cdot \alpha^2 \text{I}_{\text{probe}}) \\
- 10 \times \text{log}_{10}(\frac{1}{\beta_{\text{coh}}^2} \cdot \alpha^2 \text{I}_{\text{probe}})\\
&= 10 \times \text{log}_{10}(\frac{1}{\beta_{\text{sqz}}^2} \cdot \alpha^2 \text{I}^2_{\text{pump}}) \\
- 10 \times \text{log}_{10}(\frac{1}{\beta_{\text{coh}}^2} \cdot \alpha^2 \text{I}^2_{\text{pump}})\\
&= -20 \times \text{log}_{10}(\beta_{\text{sqz}}),
\end{align}
\textcolor{black}{which in our case are 3.36~dB and 3.44~dB calculated from the two fits in Figs.~4(a) and~(b) respectively in the main text.}

\section{Image acquisition time and spatial resolution improvement}

\textcolor{black}{As stated in the main text, acquiring the water SBS gain spectrum from 4~GHz to 6~GHz (with 40~MHz spectral resolution) in Fig.~5(b) took about 100~s, which is mainly limited by the \textit{`write'} and \textit{`read'} time of the instrument-computer interface, i.e., one needs to wait for at least 2~s for the spectrum analyzer to write the trace to its memory before the computer can view/read from it.}\\

\textcolor{black}{The quantum enhanced SBS spectra shown in Fig.~3 were obtained with 10~KHz resolution bandwidth, 10~Hz video bandwidth, and 1~s sweep time to scan a frequency span of 150~KHz (from 625~KHz to 775~KHz). In fact, one can significantly speed up the sweep time by using the \textit{`zero span'} mode of the spectrum analyzer in conjunction with reducing the resolution bandwidth or increasing the video bandwidth. In our case, the sweep time can be reduced down to 2.4~ms while the spectrum analyzer is running in the `zero span' mode with 3~KHz resolution bandwidth and 300~Hz video bandwidth.}\\

\textcolor{black}{To prove that our scheme also works with a much faster sweep time, we have retaken the SBS images of the glass triangle with the `zero span' mode aforementioned, they turned out to be essentially the same as the ones shown in Fig.~6 (but are noisier since the video bandwidth is $30\times$ wider). This implies that, our seemly slow acquisition rate is \textit{not fundamentally limited by our scheme itself}, but is rather technically limited by the instrument. This technical limitation can be readily circumvented by the use of a more advanced RF spectrum analyzer (i.e., a real-time spectrum analyzer) having a much faster data writing and read-out rate, or having a large memory that data can be processed \textit{locally}, so that the acquisition time would be solely limited by the sweep time of the spectrum analyzer. Therefore with a real-time spectrum analyzer our acquisition time would be 240~ms per 4~GHz spectrum (with 40~MHz spectral resolution), which is $120\times$ slower than the state-of-art acquisition rate \textcolor{black}{(20~ms per 4~GHz spectrum with 39~MHz spectral resolution for `real' biological tissues, in pure water or homogenous liquids/solid, it would
only require 2~ms per 4~GHz spectrum at a signal-to-noise ratio of 20~dB)} reported in Ref.~\cite{remer2020high}.  Whereas, the acquisition time in Ref.~\cite{remer2020high} was obtained with a total near-infrared excitation power of 265~mW, though they did not specify their pump and probe powers separately. In our set up the pump and probe powers are 7.5~mW and $750~\mu$W respectively, thus the total excitation power in our scheme is only 8.25~mW in the infrared (795~nm).} \\

\textcolor{black}{It is also important to point out that in addition to showing an `SNR image', it would be more useful to acquire images of practically relevant properties, such as the Brillouin frequency shifts and linewidths of the sample. This would more readily connect to real biological applications of SBS, as it is more informative of mechanical properties of the sample. In order to obtain these images, the entire spectrum must be acquired at each pixel, and then fitted for peak positions and linewidths. This, however, would require a lot longer acquisition times than the above stated.}\\

\textcolor{black}{With regard to the spatial resolution, it is very important to note that we chose to use the $f=16$~mm plano-convex lenses to focus the probe and pump beams on the sample due to concerns of loss and alignment. Objectives would be able to focus the beams much more tightly, but would also introduce much more loss that the quantum advantage would be inevitably severely degraded. A practical straightforward resolution improvement under our current situation is to use molded aspheric lenses to focus the two beams on the sample. For example, using an aspheric lens with effective focal length $f=4.5$~mm would yield a focal spot size ($1/e^2$) of $1.5~\mu$m in diameter given our beam diameter ($1/e^2$) at lens is 3~mm. Although this would improve our resolution by a factor of 3, alignment would be much more challenging as one has to overlap the pump and prove beams within an excitation volume with $1.5~\mu$m in diameter and $4.5~\mu$m in length.}\\

\section{{Measurement Sensitivity of Our Scheme}}

\textcolor{black}{Given a fixed signal amplitude, the SNR and sensitivity of measuring Brillouin peak position and linewidth are defined by the level of noise, therefore the lower the noise, the higher the sensitivity.  Our recent article~\cite{PhysRevApplied.15.044030} has a detailed discussion on this topic. Reference~\cite{remer2020high} reports the state-of-the-art performance on SBS imaging. Their acquisition time is faster than ours, due to smaller focal spot size and higher incident peak powers used for their setup; however, their sensitivity is ultimately limited by the shot-noise limit. Our scheme, however, is \textit{sub-shot-noise-limited} (our two-mode squeezed light rendered noise level is below that of the balanced coherent detection), hence our measurement sensitivity would be better than that reported in Ref.~\cite{remer2020high} if we switch to the same optical arrangement in terms of higher peak power and smaller spot size to take advantage of a higher signal due to nonlinear optical interactions, as we discussed it in great details earlier in Ref.~\cite{ballmann2015stimulated}.}

\bibliography{MyLibrary}

\end{document}